\documentclass[journal,10pt]{IEEEtran}

\usepackage{amssymb}
\usepackage{graphicx}
\usepackage{epstopdf} 
\usepackage{float}
\usepackage{subfigure}
\usepackage{verbatim}
\usepackage{multirow}
\usepackage{stfloats}
\usepackage{booktabs}
\usepackage{xcolor}
\usepackage{jsen}
\usepackage{cite}
\usepackage{amsmath,amssymb,amsfonts}
\usepackage{algorithmic}
\usepackage{textcomp}
\usepackage{wrapfig}
\def\BibTeX{{\rm B\kern-.05em{\sc i\kern-.025em b}\kern-.08em
    T\kern-.1667em\lower.7ex\hbox{E}\kern-.125emX}}
\definecolor{abstractbg}{rgb}{0.89804,0.94510,0.83137}
\begin{document}

\title{Sound Event Localization and Classification \\using Wireless Acoustic Sensor Networks in Outdoor Environments}

\author{Dongzhe Zhang, \IEEEmembership{Graduate Student Member, IEEE}, Jianfeng Chen, \IEEEmembership{Senior Member, IEEE}, \\Jisheng Bai, \IEEEmembership{Graduate Student Member, IEEE},
 Mou Wang, Dongyuan Shi, \IEEEmembership{Senior Member, IEEE},\\  Qixiang Niu, \IEEEmembership{Graduate Student Member, IEEE}, Alberto Bernardini, \IEEEmembership{Senior Member, IEEE}
\thanks{Dongzhe Zhang, Jianfeng Chen, and Jisheng Bai are with
Joint Laboratory of Environmental Sound Sensing, School of Marine Science and Technology,
Northwestern Polytechnical University, Xi’an, China (e-mail: dongzhezhang2022@mail.nwpu.edu.cn;
chenjf@nwpu.edu.cn; baijs@mail.nwpu.edu.cn).}
\thanks{Jisheng Bai and Dongyuan Shi are with the School of Electrical and Electronic Engineering,
Nanyang Technological University, Singapore (e-mail: baijs@mail.nwpu.edu.cn, dongyuan.shi@ntu.edu.sg) }
\thanks{Mou Wang is with the Institute of Acoustics, Chinese Academy of Sciences, Beijing, China (e-mail: wangmou21@mail.nwpu.edu.cn) }
\thanks{Dongzhe Zhang and Alberto Bernardini are with the Dipartimento
di Elettronica, Informazione e Bioingegneria (DEIB), Politecnico di
Milano, 20133 Milan, Italy (e-mail: dongzhezhang2022@mail.nwpu.edu.cn;  alberto.bernardini@polimi.it) }
\thanks{Qixiang Niu is with the School of Marine Science and Technology,
Northwestern Polytechnical University, Xi’an 710072, China, and also
with the Department of Signal Theory and Communications, Universidad Carlos III de Madrid, Madrid, 28903, Spain (e-mail:  niuqx@mail.nwpu.edu.cn).}
}

\IEEEtitleabstractindextext{%
\fcolorbox{abstractbg}{abstractbg}{%
\begin{minipage}{\textwidth}%
\begin{wrapfigure}[12]{r}{3in}%
\includegraphics[width=3in]{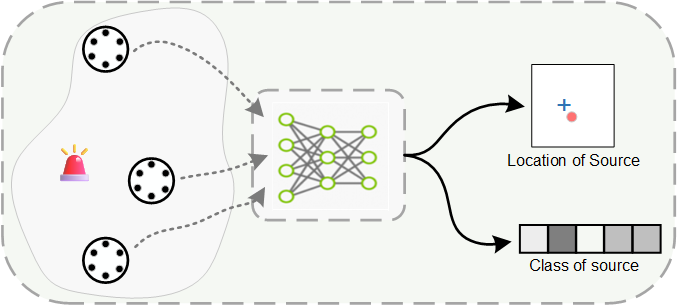}%
\end{wrapfigure}%

\begin{abstract}
The use of deep learning for sound event localization and classification with Wireless Acoustic Sensor Networks (WASNs) is an emerging research area.
However, current methods for sound event localization and classification exhibit limitations in perceiving extensive soundscapes. 
They are typically effective only for a fraction of the soundscapes and do not fully exploit the surrounding information.
Moreover, in outdoor settings, the performance accuracy is susceptible to the adverse effects of signal attenuation and environmental noise. 
In this paper, we propose a deep learning-based method that integrates frequency, temporal, and spatial domain features with attention mechanisms to estimate the location and the class of sound sources using a WASN in an outdoor setting. 
We introduce soundmap features to capture spatial information across multiple frequency bands and time frames. 
Furthermore, we integrate attention mechanisms to learn channel-wise relationships and temporal dependencies within acoustic features.
To evaluate the proposed method, we conduct experiments using simulated datasets with different levels of noise and sizes of the monitoring area, as well as different array and source positions.
Moreover, we conduct a real-world experiment in an outdoor environment with dimensions of 100 m × 80 m. 
The experimental results demonstrate the superiority of the proposed method over state-of-the-art methods in both sound event classification and sound source localization tasks. 
\end{abstract}

\end{minipage}}}

\maketitle

\section{Introduction}
Sound event localization and classification is an attractive topic and a growing research direction in the field of acoustic signal processing. 
\textcolor{black}{Although sound source localization and event classification have been extensively investigated in indoor settings \cite{yan2023nercslip}, outdoor environments present distinct conditions and additional complexities.
First, the lack of walls or ceilings means that sound propagates in a free-field manner over notably larger distances, causing high-frequency components to be absorbed more rapidly by air and often resulting in lower signal-to-noise ratios (SNRs).
Second, environmental factors such as wind, temperature, and humidity fluctuate widely and dynamically affect both the speed of sound and the degree of attenuation, making accurate modeling and real-world alignment more challenging.
Background noise in outdoor settings is typically more diverse and more non-stationary than in indoor environments. Wind rustle, bird calls, and traffic can dominate the scene, and their prevalence may change dramatically with time and location.}
To address these challenges, researchers have employed Wireless Acoustic Sensor Networks (WASNs) for outdoor sound scene monitoring\cite{faraji2019sound, huang2021robust, abeber2018distributed, liu2012distributed}. 
WASNs can integrate information from multiple sensor nodes scattered throughout the monitoring area, and maximize the system's environmental perception abilities \cite{zhang2017microphone,zhang2018rate, zhang2019distributed,zhang2022frequency }. 
Several studies have demonstrated that WASNs can facilitate the efficient monitoring of sound source activities in expansive outdoor environments and provide crucial acoustic information support for diverse application scenarios such as wildlife conservation \cite{dissanayake2018improving}, illegal intrusion detection \cite{singh2024real}, and emergency event monitoring \cite{marchegiani2022listening, 6682089}.

The Sound Event Classification (SEC) task is one of the pivotal topics in acoustic signal processing, primarily focusing on identifying specific sound sources \cite{mesaros2021sound}. 
Traditional methods for sound event classification include feature extraction and classification \cite{babaee2017overview}, template matching \cite{principi2015acoustic}, and threshold-based \cite{xia2017frame} methods. 
The Sound Source Localization (SSL) task aims to estimate the locations of sound sources, employing methods such as Direction of Arrival (DOA) \cite{kaplan2001maximum,AlbertiniICASSP2023}, Time Difference of Arrival (TDOA) \cite{canclini2012acoustic,DANG2025110488}, and Received Signal Strength Indicator (RSSI)\cite{meng2017energy}. 
These two tasks form the foundation of sound source perception, enabling a comprehensive understanding of the acoustic environment. 
\textcolor{black}{Current sound event localization and classification uses a two-stage process of classifying, then locating, or vice versa. This can propagate errors, with early inaccuracies undermining later precision.}

The continuous advancement of Deep Learning (DL) technologies has brought revolutionary changes to the field of acoustic signal processing such as noise control \cite{chiariotti2019acoustic}, automatic speech recognition  \cite{lee2016dnn}, DOA estimation \cite{nguyen2020robust, wang2022deep}, source localization \cite{feng2023soft,le2019learning} and source separation \cite{chazan2019multi}. 
The powerful nonlinear modeling capabilities of DL technologies open up the possibility of integrating the SEC and the SSL tasks. 
Sharing information between the two tasks could enhance the robustness and accuracy of the system in complex acoustic environments. 
Since 2019, the DCASE Challenge \cite{politis2020overview} has introduced indoor sound event detection and localization in Task 3, aimed at detecting and locating sound events generated in human life. 
Over the years, researchers have proposed features suitable for microphone array signals, such as SALSA \cite{nguyen2022salsa} and SALSA-lite \cite{nguyen2022salsa}. 
Paul Newman et al. \cite{marchegiani2022listening} used a multitask learning approach and signal-denoising methods to classify and locate the horns and alarms of emergency vehicles. 
\textcolor{black}{However, these two works employ only a single microphone array, yielding sound source positions solely as orientations relative to the array rather than absolute coordinates.
This limits the monitoring range, impeding effective surveillance in extensive outdoor areas.} 

Some researchers use WASNs with more than one microphone array to estimate the coordinates of target sound sources. 
Gong et al. \cite{gong2022end} proposed a DL-based end-to-end SSL method by designing a spatial-temporal model. 
This method could differentiate the global information of speakers and environments in both space and time domains. 
Ayub et al. \cite{ayub2022multiple} used histograms based on the angular distance from different nodes as association features to indicate the relationships between the frequency bins and the sources. 
Moing et al. \cite{le2019learning} proposed a DL-based end-to-end method to model the mapping between the locations of multiple sources and the multi-channel short-time Fourier transform (STFT) features of the arrays. 
They expanded the work and proposed the use of adversarial learning to close the gap between synthetic domains and real domains. 
Kindt et al. \cite{kindt20212d} proposed decentralized deep neural networks to allow different arrays to collaborate effectively.
However, these methods only model the sound signal in the temporal domain and in the frequency domain, without fully considering the spatial features of the sound sources. 
\textcolor{black}{The methods in \cite{le2019learning, feng2023soft} rely on general-purpose time-frequency features, such as the real and imaginary parts of the STFT. 
From these features, they implicitly learn spatial information (e.g., inter-channel phase and amplitude differences). 
Furthermore, these approaches focus on the SSL single task while overlooking the SEC task.}

\textcolor{black}{WASN expands the monitoring coverage beyond the limited range of a single array, thus enabling large-scale, comprehensive observations. 
Moreover, as each node captures the source signals from a slightly different point, the multi-view information across nodes helps improve both localization accuracy and event classification performance. 
In particular, while one node may be significantly affected by noise or reverberation, the others can still provide reliable information, thereby mitigating potential misclassifications.}
In this paper, we introduce a novel DL-based method for the sound event localization and classification task. 
We use a WASN for signal acquisition and processing, which includes multiple microphone arrays.
Our proposed method incorporates features that integrate information across frequency, temporal, and spatial domains, providing a comprehensive understanding of complex sound scenes. 
Inspired by \cite{dibiase2000high}, we propose the soundmap feature, which is based on multiple frequency sub-bands and represents the spatial energy distribution of various sound sources in the time-frequency domain. 
The soundmap feature uses geometric information from the array to enhance the spatial gain of the sources while suppressing noise interference, thus enabling effective extraction of spatial information in low signal-to-noise ratio (SNR) outdoor environments. 
Then, we use Gammatonegram \cite{holdsworth1988implementing} to represent the sound signals received by each array. 
Gammatonegram is formed by feeding the sound signals into a gammatone filter bank. 
It aligns well with human auditory characteristics and has been proven to be more effective than Mel-Frequency Cepstrum Coefficients (MFCCs) \cite{zhuang2008feature} in outdoor settings \cite{marchegiani2022listening}. 
Furthermore, we introduce a multitask model based on Convolutional Neural Networks (CNNs) and Transformer encoder modules. 
Specifically, CNNs extract local invariant features, while the multi-head self-attention mechanism \cite{vaswani2017attention} is employed to learn the significance of channel-wise features. 
By employing different loss functions for backpropagation, the model effectively integrates the SEC and SSL task characteristics. 
Finally, we evaluate the proposed method in various acoustic environments and real-world tests. 
\textcolor{black}{To ensure robustness to arbitrary network geometries, our model incorporates the spatial coordinates of the sensor arrays as an input feature and is validated on deployment configurations unseen during training.}
Experimental results demonstrate that our method outperforms state-of-the-art methods across different noise levels and interference sources, as well as different arrays and source positions. 

The rest of the paper is structured as follows: 
Section II focuses on the proposed features and the multitask model. 
Section III describes the datasets, the experimental setup, and the evaluation metrics. 
The experiments and performance of the system are discussed in Section IV. 
Finally, Section V concludes this work.

\begin{figure}[h]
\centering
\includegraphics[scale=0.7]{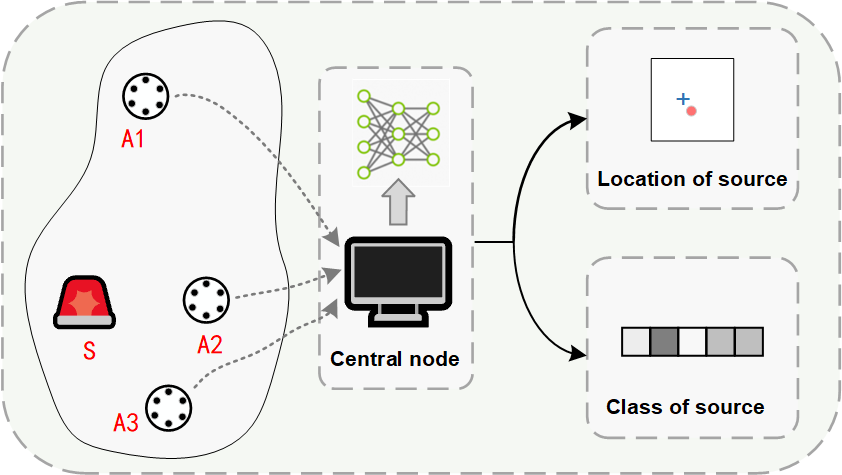} \hspace{-5mm}
\caption{WASN system where A1, A2, and A3 represent array nodes, while S represents the target sound source. Array nodes can collect and process multi-channel audio signals, extract various features from the signals, and transmit these features to the central node. The central node collects feature data from multiple array nodes and uses a neural network to estimate the class and locations of the target source.}
\label{fig:structure}
\end{figure}

\section{PROPOSED METHOD}

\textcolor{black}{Let us consider a WASN system in an outdoor area as the one shown in Fig. \ref{fig:structure}, consisting of $N$ nodes and a central node, where each node is a microphone array with $M$ microphone sensors arranged in a uniform circular geometry. 
During the operation of the WASN system, each array node processes the multi-channel acoustic signal collected by the $M$ microphone sensors and then sends the results to the central node. 
The central node is responsible for receiving and processing information from each array node and estimating the class and coordinates of the target sound source by using a neural network-based algorithm.
We use three types of feature to train the neural network: soundmap features, Gammatonegram features, and coordinates of array nodes. 
Then, we employ CNNs and attention mechanisms to learn the mapping relationship between features and the position, as well as the class of the sound source.
The following subsections elaborate on the main components of the proposed method.
}

\subsection{Features}

\subsubsection{Soundmap features}
\textcolor{black}{Broadband beamforming \cite{high1993low} is a commonly used method in array signal processing, which allows microphone arrays to perceive spatial information within specific frequency bands. 
Assuming that the array node has sufficient data processing and storage capabilities, it can perform a beamforming algorithm to obtain soundmap features. 
This study focuses on a 2D model, where sound sources and microphone arrays are assumed to be in the same plane.}

\textcolor{black}{For each array node, the continuous time signal received by the $m$th microphone sensor can be expressed as:
\begin{equation}
   x_{m}(t)=\sum_{k=1}^{K} s_{k}(t) * h_{m}\left(t, \theta_{k}\right)+v_{m}(t) \quad m=1,2, \cdots, M\label{equation:1}
\end{equation} 
where $K$ is the total number of sources, $t$ represents the time variable in seconds, $s_k$ represents the signal from the $k$th source, $h_{m}\left(t, \theta_{k}\right)$ denotes the impulse response from the position of the $k$th source located at direction $\theta_k$ to the $m$th sensor, and $v_m(t)$ represents the additive noise of the $m$th sensor. The steered response power of the broadband beamformer can
be expressed as follows:
\begin{equation}
P(\theta, f)=\sum_{u=1}^{M} \sum_{v=u+1}^{M}\psi_{u v}(f) X_{u}(f) X_{v}^{*}(f) e^{j 2\pi f(\tau(\theta, u)-\tau(\theta, v))}
\end{equation}
where $f$ is the frequency variable in Hertz, $X_m(f)$ denotes the Fourier transform of $x_m(t)$ with $X^{*}_m(f)$ its complex conjugate, $\tau(\theta,u)$ is the time delay associated with the direction $\theta$ relative to the microphone sensor with index $u$. 
\textcolor{black}{The time-delay model for $\tau(\theta, u)$ is based on the far-field plane-wave propagation assumption.
This assumption holds when the sound source is sufficiently distant from the array, causing the incoming wavefronts to be approximately planar.
Given our large-scale outdoor monitoring scenarios where sources are typically tens to hundreds of meters away from the arrays, this condition is well-satisfied.}
The term $\psi_{u v}(f)$ is a weighting function, which for the SRP-PHAT method \cite{high1993low} is set as:
\begin{equation}
\psi_{u v}(f)=\frac{1}{\left|X_{u}(f) X_{v}^{*}(f)\right|} \,\,.
\end{equation} 
When performing spatial scanning, the target frequency band is chosen using a range [$f_1$, $f_2$], where $f_1$ is the starting frequency and $f_2$ is the ending frequency. The output of the broadband beamforming can be represented as:
\begin{equation}
\bar{P}(\theta)=\int_{f_1}^{f_2} P(\theta, f) \, df
\end{equation}}

\textcolor{black}{By scanning spatial directions in the range $\left[\theta_a, \theta_b \right]$, we obtain spatial-energy information around the array, which we refer to as a soundmap feature. This feature can be represented as a vector of power values at different directions: $\bar{\mathbf{p}}=[\bar{P}(\theta_a),\dots, \bar{P}(\theta_b)]$. 
When using broadband beamforming to scan the space, signals from a specific direction are amplified while signals from all other directions are suppressed. 
Furthermore, due to the unique power distributions across the frequency spectrum of different classes of sound samples, the soundmap feature can illustrate the frequency-domain distribution of the target signal \cite{zhang2024multiple}. 
We partition the frequency spectrum into $F$ sub-bands and perform broadband beamforming across $U$ distinct angles for each sub-band.
In a setup with $N$ array nodes, the soundmap features extracted at the central node have dimensions of $N$×$F$×$U$.}

\begin{figure*}[b]
\begin {center}
\includegraphics[width=0.99\textwidth]{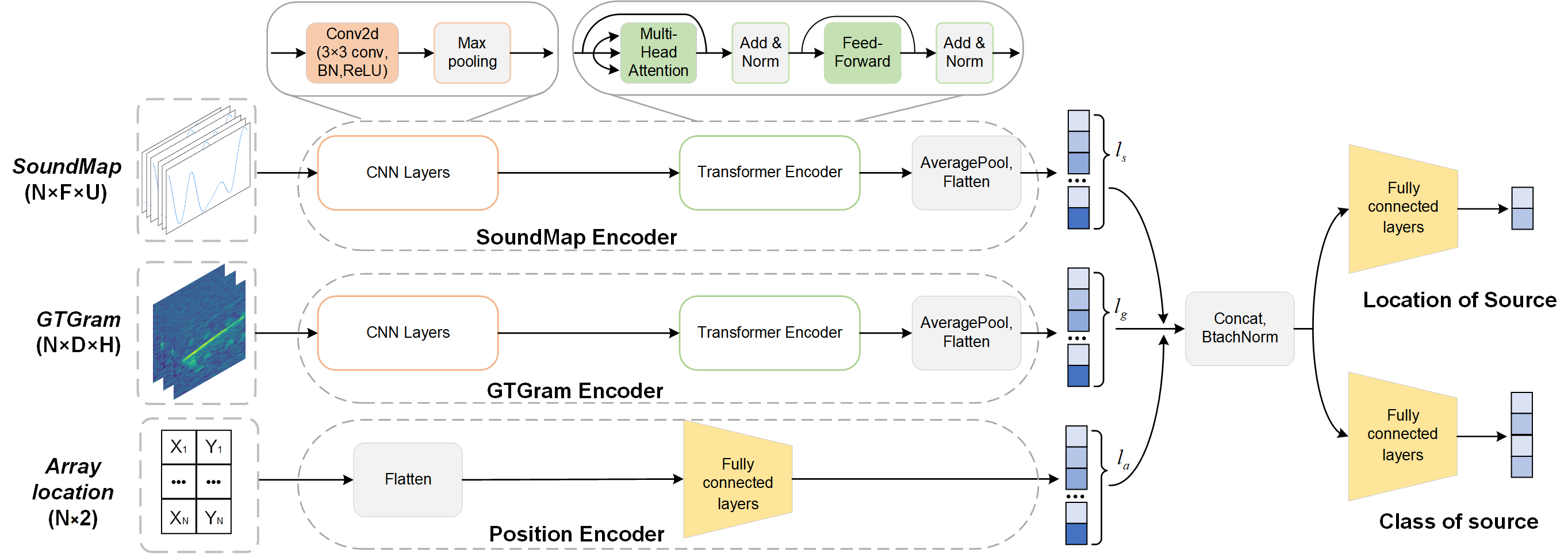}
\caption{The model architecture of the proposed method.}
\label{fig:network}
\end {center}
\end{figure*}

\subsubsection{Gammatonegram (GTGram) features}

\textcolor{black}{MFCCs are widely used in audio tasks. However, recent research \cite{chakrabarty2016abnormal} shows that MFCCs have limitations in specific acoustic environments, especially those with high levels of noise and dynamic conditions like traffic scenes and outdoor sound source monitoring. 
In contrast, Gammatone \cite{holdsworth1988implementing} representations are highly effective in various audio classification tasks, even in the presence of strong interference and noise. 
Therefore, we use Gammatone filterbanks to generate GTGram features. These filterbanks, originally designed to approximate the human cochlear frequency response, provide a perceptually relevant representation of audio signals. 
The impulse response of a Gammatone filter can be expressed as:
\begin{equation}
g(t) = t^{(n-1)}e^{-2\pi bt} \cos(2\pi f_c t)
\end{equation} 
where $t$ represents the time variable in seconds, $n$ represents the filter order, $b$ represents the bandwidth of the filter, and $f_c$ represents the center frequency in Hertz. 
\textcolor{black}{We employ 4th-order gammatone filters ($n=4$). The center frequencies $f_c$ of each band are spaced from 20 Hz to 4000 Hz according to the Equivalent Rectangular Bandwidth (ERB) scale \cite{glasberg1990derivation, toshio1995optimal}. 
The bandwidth $b$ of each filter is also determined by its center frequency based on the ERB model by using the equation: $b = 1.019 \times (24.7 + 0.108 \times f_c)$.}
When generating GTGram features, we divide the audio into $D$ frames in the time domain and partition it into $H$ segments in the frequency domain. 
In a setup with $N$ array nodes, the GTGram features have dimensions of $N$$\times$$D$$\times$$H$. }

\subsubsection{Coordinates of array nodes}
\textcolor{black}{We consider the variability of the deployment of array nodes in the WASN system, acknowledging that their positions may change with different usage scenarios. 
So we incorporate the positions of the array nodes as inputs to the neural network, and the input size of the position encoder is $N \times 2$, representing the normalized 2D coordinates \cite{diaz2020robust} of the array nodes. }

\begin{table*}[]
\centering
\caption{Detailed architecture and parameter settings of the proposed method}

\begin{tabular}{clclcl}
\hline
\multicolumn{2}{c|}{\textbf{Soundmap Encoder}}                                                       & \multicolumn{2}{c|}{\textbf{GTGram Encoder}}                                                             & \multicolumn{2}{c}{\textbf{Position Encoder}}             \\ \hline
\multicolumn{2}{c|}{\begin{tabular}[c]{@{}c@{}}Conv 3×3 @ 64\\ BN, ReLU\end{tabular}}       & \multicolumn{2}{c|}{\begin{tabular}[c]{@{}c@{}}Conv 3×3 @ 64\\ BN, ReLU\end{tabular}}       & \multicolumn{2}{c}{\multirow{5}{*}{\begin{tabular}[c]{@{}c@{}}Flatten\\ FC 512, ReLU\end{tabular}}} \\ \cline{1-4}
\multicolumn{2}{c|}{Max pooling 1×3}                                                         & \multicolumn{2}{c|}{Max pooling 3×3}                                                         & \multicolumn{2}{c}{}                             \\ \cline{1-4}
\multicolumn{2}{c|}{\begin{tabular}[c]{@{}c@{}}Conv 3×3 @ 128\\ BN, ReLU\end{tabular}}      & \multicolumn{2}{c|}{\begin{tabular}[c]{@{}c@{}}Conv 3×3 @ 128\\ BN, ReLU\end{tabular}}     & \multicolumn{2}{c}{}                             \\ \cline{1-4}
\multicolumn{2}{c|}{Max pooling 1×3}                                                         & \multicolumn{2}{c|}{Max pooling 3×3}                                                         & \multicolumn{2}{c}{}                             \\ \cline{1-4}
\multicolumn{2}{c|}{\begin{tabular}[c]{@{}c@{}}Conv 3×3 @ 128\\ BN, ReLU\end{tabular}}      & \multicolumn{2}{c|}{\begin{tabular}[c]{@{}c@{}}Conv 3×3 @ 128\\ BN, ReLU\end{tabular}}      & \multicolumn{2}{c}{}                             \\ \hline
\multicolumn{2}{c|}{Max pooling 1×3}                                                         & \multicolumn{2}{c|}{Max pooling 2×2}                                                         & \multicolumn{2}{c}{\multirow{5}{*}{FC 90, ReLU}}  \\ \cline{1-4}
\multicolumn{2}{c|}{Transformer encoder  ×2}                                               & \multicolumn{2}{c|}{Transformer encoder  ×2}                                               & \multicolumn{2}{c}{}                             \\ \cline{1-4}
\multicolumn{2}{c|}{\begin{tabular}[c]{@{}c@{}}Average pooling 32 × 32\\ Flatten\end{tabular}} & \multicolumn{2}{c|}{\begin{tabular}[c]{@{}c@{}}Average pooling 16 × 16\\ Flatten\end{tabular}} & \multicolumn{2}{c}{}                             \\ \cline{1-4}
\multicolumn{2}{c|}{FC 512, ReLU}                                                            & \multicolumn{2}{c|}{FC 128, ReLU}                                                            & \multicolumn{2}{c}{}                             \\ \cline{1-4}
\multicolumn{2}{c|}{FC 60, ReLU}                                                             & \multicolumn{2}{c|}{FC 60, ReLU}                                                             & \multicolumn{2}{c}{}                             \\ \hline
\multicolumn{6}{c}{\begin{tabular}[c]{@{}c@{}}Concate,  BN\end{tabular}}                                                                                                                                                                     \\ \hline
\multicolumn{2}{c|}{\multirow{2}{*}{\textbf{Output}}}                                           & \multicolumn{2}{c|}{FC 512, ReLU}                                                            & \multicolumn{2}{c}{FC 512, ReLU}                  \\ \cline{3-6} 
\multicolumn{2}{c|}{}                                                                       & \multicolumn{2}{c|}{FC $I$}                                                            & \multicolumn{2}{c}{FC 2 }                  \\ \hline
\end{tabular}
\label{tab_network}
\end{table*}

\subsection{Deep neural network architecture}

\textcolor{black}{Given soundmap features, GTGram features, and coordinates of array nodes, a DL model is used to learn the mapping between these features and the location, as well as the class of the sound source. 
Figure \ref{fig:network} and Table \ref{tab_network} illustrate the proposed network model. }

\textcolor{black}{CNNs were originally developed in the literature for tasks related to image processing, primarily image classification. 
The applications have expanded to audio-related tasks such as speech recognition \cite{abdel2014convolutional} and DOA estimation \cite{nguyen2020robust} in recent years. 
In the proposed architecture, the features are passed through convolutional layers, which consist of a set of trainable kernels. 
By spanning all the channels, these kernels enable the convolutional layer to learn relevant inter-channel features, thereby enhancing the model's ability to capture complex patterns of the input features. 
Moreover, the use of kernels across all channels allows for the extraction of local invariant features from the input features. 
As far as the soundmap features are concerned, kernels can learn patterns related to frequency bands and spatial range. 
As far as the GTGram features are concerned, kernels can learn patterns related to frequency and frame. }

\textcolor{black}{To further enhance the training process, batch normalization \cite{ioffe2015batch}, maxpooling, and Rectified Linear Unit (ReLU) activation functions \cite{nair2010rectified} are used. 
Batch normalization helps accelerate training by normalizing the input values to each layer, while ReLUs introduce nonlinearity to the network, enhancing the model's ability to learn complex relationships within the data.
The output of convolutional layers are feature maps. Maxpooling downsamples feature maps from the convolutional layers, preserving the most important features and reducing computational complexity.}

{\color{black}{
To model sequential dependencies and correlations among input features, the Transformer \cite{vaswani2017attention} is incorporated into the model. 
The Transformer encoder leverages the Multi-Head Self-Attention (MHSA) mechanism to compute relationships among time steps and spatial angles in the feature sequence. For a given input feature $\boldsymbol{R}$, which represents the output of the preceding neural network layer, the Multi-Head Self-Attention (MHSA) mechanism transforms the input into query, key, and value representations to compute attention scores. These scores are normalized and combined to produce the multi-head output $\boldsymbol{O}$:
\begin{equation}
\boldsymbol{O} = \operatorname{Concat}\left( \boldsymbol{h}_1, \boldsymbol{h}_2, \cdots, \boldsymbol{h}_h \right) \boldsymbol{W}^{O},
\end{equation}
where $\operatorname{Concat}(\cdot)$ denotes the concatenation of multiple vectors along the feature dimension, $\boldsymbol{W}^{O}$ is a learnable weight matrix that projects the concatenated output into the desired feature space, and each vector $\boldsymbol{h}_i$ (for $i = 1, 2, \dots, h$) is the output of the $i$-th attention head, which captures the weighted contributions of different sequence elements. 
Feed-Forward Network (FFN), residual connections, and Layer Normalization (LN) \cite{ba2016layer} are applied to enhance gradient flow and training stability:
\begin{equation}
\boldsymbol{O}^{\prime} = \mathrm{LN}\left(\mathrm{FFN}\left(\mathrm{LN}(\boldsymbol{R} + \boldsymbol{O})\right) + \mathrm{LN}(\boldsymbol{R} + \boldsymbol{O})\right) \,.
\end{equation}
The output $\boldsymbol{O}^{\prime}$  of the Transformer encoder captures the temporal and spatial dependencies of the input features, which are essential for the accurate localization and classification of sound sources.
By integrating CNNs for feature extraction and the Transformer encoder for sequence modeling, the proposed model effectively captures both local and global patterns in the input data.

In Fig. \ref{fig:network}, the soundmap encoder outputs a vector of length $l_s$, the GTGram encoder outputs a vector of length $l_g$, and the position encoder outputs a vector of length $l_p$. 
\textcolor{black}{These vectors are concatenated, and batch normalization is applied to ensure consistent value distributions. 
Following this, as detailed in Table I, several fully connected layers are used to process the combined features and produce the final outputs for the two tasks.}
For the SSL task, the fully connected layers output a two-dimensional vector, representing the 2D coordinates of the target sound source. 
Since the input of the position encoder is normalized, predictions should be multiplied by the area size to obtain the estimated source coordinates. 
For the SEC task, the fully connected layers output a vector of length $I$, representing predictions for $I$ classes of sound events. 

\subsection{Loss function}
Mean Squared Error (MSE) is used as the loss function for SSL, represented as $L_1$. 
Binary Cross Entropy (BCE) is used as the loss function for SEC, denoted as $L_2$.  The overall loss function can therefore be expressed as $L = L_1 + \lambda \cdot L_2$.}
\textcolor{black}{To balance the contributions of the two loss components and avoid numerical imbalances during training, we introduce a weight $\lambda$ for $L_{2}$. 
The value of $\lambda$ is determined empirically through a grid search over a range of values \{0.001, 0.01, 0.1, 1.0, 10, 100\} on our validation set. 
We find that $\lambda=0.1$ provides the best trade-off, leading to robust convergence and optimal combined performance on both tasks. 
This weighting ensures that the MSE loss for localization and the BCE loss for classification contribute comparably to the total gradient, facilitating effective multitask learning.}

\section{EVALUATION}
In this section, we introduce simulated multi-array datasets, evaluation metrics of the proposed method, and baseline methods.
\subsection{Datasets}

\textcolor{black}{To achieve both geometric flexibility and physical realism in our large-scale outdoor simulation, we adopt a two-stage hybrid simulation framework. 
In the first stage, we utilize the Pyroomacoustics package \cite{scheibler2018pyroomacoustics} for geometric and temporal modeling. 
By setting the image source model order to zero (\texttt{max\_order=0}), we created an anechoic environment. 
This allows us to efficiently configure complex spatial layouts of sources and microphone arrays and to calculate the direct-path time-of-flight.
In the second stage, the signals are processed with a custom physics-based attenuation filter. 
For each source-microphone pair, we use a frequency-domain filter to model attenuation based on the international standard ISO 9613-2 \cite{international1993acoustics}. 
This filter incorporated two effects:}
\begin{enumerate}
    \item \textcolor{black}{Frequency-independent geometrical divergence ($A_{\text{div}}$), which models the spherical spreading of the sound wave. This attenuation is calculated according to the formula:}
    \begin{equation}
        \textcolor{black}{A_{\text{div}} = 20 \cdot \mathrm{log}_{10}(d/d_0) + 11 \quad (\text{dB})}
    \end{equation}
    
    \textcolor{black}{where $d$ is the source-to-receiver distance and $d_0$ is the reference distance (1 m).}

    \item \textcolor{black}{Frequency-dependent atmospheric absorption ($A_{\text{atm}}$), which accounts for the absorption of sound energy by the air. This attenuation is calculated for each frequency band $f$ using the formula:}
    \begin{equation}
       \textcolor{black}{ A_{\text{atm},f} = \frac{\alpha_f \cdot d}{1000} \quad (\text{dB})}
    \end{equation}
    \textcolor{black}{where $d$ is the distance in meters, and $\alpha_f$ is the atmospheric attenuation coefficient in dB/km for the specific frequency band $f$.}
\end{enumerate}

\textcolor{black}{The filter's frequency response is shaped according to the total attenuation $A_f = A_{div} + A_{atm,f}$. 
 We use the specific attenuation coefficients ($\alpha_f$) from \cite{international1993acoustics} for our stated conditions ($20^{\circ}$C, 70\% humidity) to ensure that higher frequency components are attenuated more significantly over distance, accurately reflecting real-world outdoor propagation.}

\textcolor{black}{Table~\ref{tab1} summarizes the experimental parameters. Each microphone array node consists of a circular 8-element microphone array with a radius of 11cm, and the sampling frequency is set to 8000 Hz. We designate the first microphone of each array node as the reference microphone. Pyroomacoustics then simulates the attenuation of these multi-channel signals based on their physical properties (e.g., the decline in acoustic intensity due to the enhanced air absorption setting), ensuring that high-frequency content is attenuated more rapidly with increasing propagation distance. This setup closely emulates realistic acoustic behavior in large outdoor areas while taking advantage of an efficient simulation toolkit.}

We consider three types of sound events: emergency siren, human scream, and gunshot. 
To enhance the robustness of the system, we introduce a noise category, which represents the absence of a sound source or the presence of only interfering sources. 
\textcolor{black}{Therefore, with four sound event classes ($I$=4), the SEC task can be viewed as a multi-class classification task.}
To ensure that the simulated data closely approximate real-world scenarios, we prepare multiple samples for each sound class during the data generation phase. 
For each generated sample, at most one target source and two interfering noise sources are active simultaneously. 
For the target source, we extract the samples from publicly available databases such as UrbanSound8K \cite{salamon2014dataset} and www.freesound.org. 
We initially apply voice activity detection \cite{sohn1999statistical} to filter out inactive portions and select segments containing the desired signal, followed by segmenting the samples into 1-second audio clips. 
There are 35 minutes of siren recordings, 46 minutes of scream recordings, and 37 minutes of gunshot recordings. 
For interference sources, we used the CAS dataset \cite{bai2024description}, comprising environmental sounds recorded in 12 cities in China. 
Specifically, we select the \textit{Urban park} category from the CAS dataset, which includes urban park environmental sounds such as children playing, birdcall, and dog barking. 
We select 50 minutes of recordings from this category for data generation.

We assign different ranges of Sound Pressure Level (SPL) for each sound event. 
We assume a temperature of 20°C and humidity of 70\%.  We simulate the attenuation of sound as it propagates through spherical diffusion and undergoes atmospheric absorption. 
Therefore, the SPL and the SNR measured for each array are related to the propagation distance. 
For the simulated data, we generate five areas with varying sizes. 
The sizes of these areas in the simulated dataset are listed in Table \ref{tab2}. 
We assume that all the array nodes and the sources are deployed at a consistent height of 1.5 \text{m} from the ground level. 
For each simulated sample, we begin by randomly selecting an area from Table \ref{tab2}. 
Subsequently, we randomly choose five array positions on a grid with a resolution of $1~\text{m} \times 1~\text{m}$, ensuring a minimum distance of 30 \text{m} between any two arrays. 
Then, we randomly designate the source positions on a grid with a resolution of $1~\text{m} \times 1~\text{m}$ and synthesize recordings by generating 1-second audio samples.
In training, validation and test sets, the positions of arrays and sources do not overlap, with each set containing 90k, 25k, and 15k audio samples, respectively.

\begin{table}[t]
	\centering
	\caption{Parameters for generating the dataset}
	\begin{tabular}{lc}
		\toprule  

		Parameter&Value \\  
		\midrule  
            Node type     &     8-element UCA\\
		Array  radius     &     11 cm \\
            Number of nodes     &      5 \\
            Sampling frequency     &     8000 Hz \\
            Length of sample    &   1 s \\
            SPL  of  siren    &     {[100, 120] (dB)} \\
            SPL  of  scream    &    {[90, 110] (dB)} \\
            SPL  of  gunshot   &   {[120, 140] (dB)} \\
            SPL  of  interfering source   &    {[90, 130] (dB)} \\
            SPL  of background  noise   &  {[40, 70] (dB)} \\ 
            \textcolor{black}{Temperature}   &     \textcolor{black}{$20^{\circ}$C}\\
		  \textcolor{black}{Humidity}     &      \textcolor{black}{70\%}\\            
            \bottomrule  
	\end{tabular}
        \label{tab1}
\end{table}

\begin{table}[h]
\centering
\caption{\textcolor{black}{Area size in dataset}}
\linespread{0.2}
\begin{tabular}{ccc}
    \toprule  
    Dataset                 & AreaID & Size (\text{m}) \\ \hline
    \multirow{5}{*}{\begin{tabular}[c]{@{}c@{}}Training set \&\\ validation set\end{tabular}} & Area 1  &  $100 \times 100$   \\ 
                            & Area 2  &  $100 \times 180$   \\
                            & Area 3  &  $120 \times 120$  \\
                            & Area 4  &  $160 \times 180$   \\
                            & Area 5  &  $200 \times 200$   \\  \hline
    \multirow{3}{*}{Test set}   & Area 6 & $140 \times 140$   \\
                            & Area 7 & $140 \times 180$  \\
                            & Area 8 & $180 \times 180$  \\
    \bottomrule  
    \label{tab2}
\end{tabular}
\end{table}

\subsection{Evaluation metrics}
The classification output is a $I$-dimensional vector representing the probability distribution across the $I$ classes of sound sources. 
To assess the performance of classification, we consider four metrics: precision (\text{Pre}), recall, F1 score, and False Alarm Rate (\text{FAR}) \cite{donohue2011constant}. For each class $c \in \{1, \ldots, I\}$, the metrics are defined as follows:
\begin{equation}
\mathrm{Pre}_c = \frac{\mathrm{TP}_c}{\mathrm{TP}_c + \mathrm{FP}_c}
\end{equation}
\begin{equation}
\mathrm{Recall}_c = \frac{\mathrm{TP}_c}{\mathrm{TP}_c + \mathrm{FN}_c}
\end{equation}
\begin{equation}
\mathrm{F1}_c = 2 \times \frac{\mathrm{Pre}_c \times \mathrm{Recall}_c}{\mathrm{Pre}_c + \mathrm{Recall}_c}
\end{equation}
where $\mathrm{TP}_c$ represents true positives, $\mathrm{FP}_c$ represents false positives, and $\mathrm{FN}_c$ represents false negatives. 
In addition, we pay attention to the false alarm of the model. We consider interfering sources as non-active classes and other classes as active classes such that \cite{donohue2011constant}
\begin{equation}
\mathrm{FAR} =  \frac{\mathrm{FP}_\mathrm{all}}{\mathrm{TOT}_{\mathrm{nonAct}}}
\end{equation}
where $\mathrm{FP}_\mathrm{all}$ represents the number of data of non-active classes incorrectly classified as belonging to an active class, while $\mathrm{TOT}_{\mathrm{nonAct}}$ is the total number of data belonging to non-active classes.

For the SSL task, we use the Root Mean Square Error (\text{RMSE}) as a metric to measure the performance of different methods:

\begin{equation}
\mathrm{RMSE} = \sqrt{(x_p - x_t)^2 + (y_p - y_t)^2}
\end{equation}
where $x_p$ and $y_p$ are the predicted coordinates, and $x_t$ and $y_t$ are the ground-truth coordinates of the target sound source.

We also design a comprehensive metric to compare the performance of different sound event localization and classification methods:
\begin{equation}
\mathrm{SELC}_{\mathrm{score}} = \frac{\left(\mathrm{F1} + \left(1 - \mathrm{FAR}\right) + \left(1 - \frac{\mathrm{RMSE}}{\mathrm{len}_{\mathrm{area}}}\right) \right)}{3}
\end{equation}
where \text{F1} represents the mean F1 score of each class, and $\mathrm{len}_{\mathrm{area}}$ denotes the diagonal length of the testing area. 
\textcolor{black}{We design the $SELC_{score}$ to provide an overall and balanced metric that reflects the overall practical utility of a system for our target applications, such as public safety and security monitoring. 
The equal weighting of its three components is a deliberate choice rooted in this application-driven perspective.}

\subsection{Implementation details}

For the soundmap features, we consider a frequency range from 20~Hz to 4000~Hz, which is then divided into 6 sub-bands.
The spatial range is steered from 0° to 359°. 
For the GTGram features, a gammatone filterbank with 64 frequency channels is used. 
Filtering is applied to the time domain with frames of 100~ms duration with 50~ms overlap, using a Hamming window to reduce spectral leakage.
For the array position features, the 2D coordinates of each array are normalized to accommodate varying area sizes.
\textcolor{black}{The network is trained using the Adam optimizer with an initial learning rate of $1 \times 10^{-4}$. We employ a step decay learning rate schedule, reducing the learning rate by a factor of 0.1 every 15 epochs to facilitate stable convergence.
The model is trained for a maximum of 50 epochs with a batch size of 32. 
To prevent overfitting and select the best performing model, we use an early stopping mechanism with a patience of 5 epochs, monitoring the total loss on the validation set. 
All models are implemented using the PyTorch framework and trained on an NVIDIA GeForce RTX 3090. }
\textcolor{black}{The model's deployment is based on a distributed edge-computing architecture.
Feature extraction is performed on the individual array nodes, and the central node is responsible for collecting features and performing the final inference. 
The detailed hardware architecture and performance metrics of this edge-computing implementation are presented in Section IV.D.}

\subsection{Baseline methods}
We introduce four baseline methods for comparison with our proposed approach in the SSL task and three baseline methods in the SEC task. 
To ensure a fair comparison across different input features and neural networks, all methods are carefully fine-tuned to our simulated data.\\
\textbf{SEC-CNN} \cite{mushtaq2020environmental}: It uses deep CNNs and data augmentation techniques with audio features such as Mel spectrogram, MFCC, and Log-Mel. We use the waveform from the first channel of each array as the raw data to generate features.\\
\textbf{SEC-UNET} \cite{marchegiani2022listening}: It focuses only on the classification branch and employs hard parameter sharing and the Unet network. The data from the first channel of each array is used as the raw input to generate features.\\
\textcolor{black}{\textbf{SEC-CRNN} \cite{bansal2024robust}:  It employs a Convolutional Recurrent Neural Network (CRNN) that leverages CNN layers to extract local time-frequency features from the waveform of the first channel, and  Long Short Term Memory (LSTM) to capture temporal dynamics. }   \\
\textbf{SSL-PLSE} \cite{koks2001passive}: It uses DOA estimation and direction cosine intersection method to obtain the position of the sound source. Since each array introduces errors in DOA estimation, when multiple arrays perform direction cosine intersection, the position of the sound source is estimated within a certain area. Therefore, the pseudo-linear least squares method (PLSE) is needed to improve the localization accuracy.\\
\textbf{SSL-FUZZY} \cite{faraji2019sound}: It firstly defines and quantifies a spatial Region of Interest (ROI). Then, the fuzzy belief value of sound source presence is estimated. Finally, a defuzzification process is applied to determine the precise location of the sound source. Consistent with the baseline method, we employ the triangular fuzzifier, the product t-conorm, and the maximum-based defuzzifier to obtain the coordinates of the source.\\
\textbf{SSL-STFT} \cite{le2019learning}: It uses both the real and imaginary components of STFT features from the arrays, which are integrated into an encoding-decoding architecture. This method uses a Heat Map representation (HM-rep) and array-encoder architecture. Additionally, we refine the neural network structure and incorporate array locations as input features during the training stage.\\
\textcolor{black}{\textbf{SSL-SOFT} \cite{feng2023soft}: It uses a soft-label encoding strategy, converting each ground-truth source position into a probability distribution rather than a single point. Interpreting the source location as a distribution makes it less vulnerable to minor misalignments. In practice, we encode each array node’s coordinates alongside time-frequency features, feeding both into a neural network trained with a distribution-focused objective. }

\section{RESULTS AND DISCUSSION}
In this section, we assess the baseline methods and the proposed method and provide further discussions. 
We first compare the performance of the proposed method and the baseline methods for both SEC and SSL tasks. 
We then analyze the factors that contribute to the errors. 
Finally, we conduct real-world experiments to validate the practicality of the system. 

\begin{table}[b]
\centering
\caption{\textcolor{black}{SEC performance of different methods}}
\label{tableVA}
\begin{tabular}{cccccc}
\hline
\multirow{2}{*}{Metrics} & \multicolumn{4}{c}{$\mathrm{F1}$  score $\uparrow$}                                      & \multirow{2}{*}{$\mathrm{FAR}$ $\downarrow$} \\ \cline{2-5}
                         & siren          & scream         & gunshot       & interfering    &                      \\ \hline
SEC-CNN\cite{mushtaq2020environmental}                  & .941          & .908          & .974          & .945          & .064                \\
SEC-UNET\cite{marchegiani2022listening}                  & .930          & .899          & .981          & .927          & .072                \\
SEC-CRNN\cite{bansal2024robust}                  & .948         & .908          & .979          & .947          & .061                \\

\multirow{2}{*}{\begin{tabular}[c]{@{}c@{}}proposed \\ (only $SEC$)\end{tabular}} & \multirow{2}{*}{.933} & \multirow{2}{*}{.882} & \multirow{2}{*}{.978} & \multirow{2}{*}{.932} & \multirow{2}{*}{.063} \\ \\

\multirow{2}{*}{\begin{tabular}[c]{@{}c@{}}proposed \\ (w/o attention)\end{tabular}} & \multirow{2}{*}{.935} & \multirow{2}{*}{.884} & \multirow{2}{*}{.950} & \multirow{2}{*}{.879} & \multirow{2}{*}{.067} \\ \\

\textbf{proposed}           & \textbf{.963} & \textbf{.911} & \textbf{.989} & \textbf{.952} & \textbf{.039}       
\\\hline
\end{tabular}
\end{table}

\subsection{Sound event classification}

\begin{figure}[htbp]
    \centering
    \subfigure[Confusion matrix without attention]{
        \includegraphics[width=0.45\textwidth]{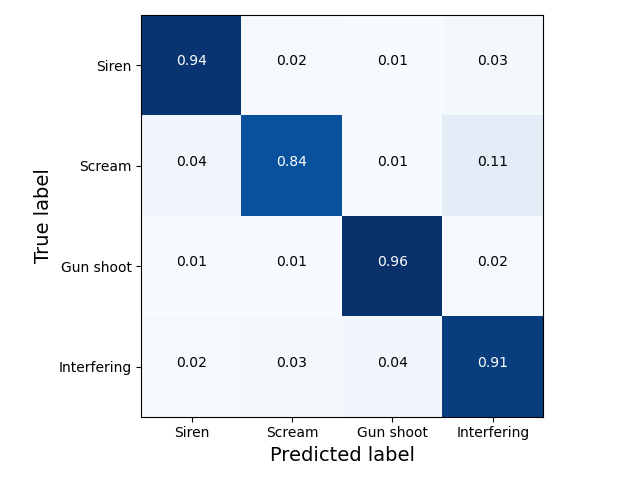}
        \label{subfig:cm_no_att}
    }
    \subfigure[Confusion matrix with attention (proposed method)]{
        \includegraphics[width=0.45\textwidth]{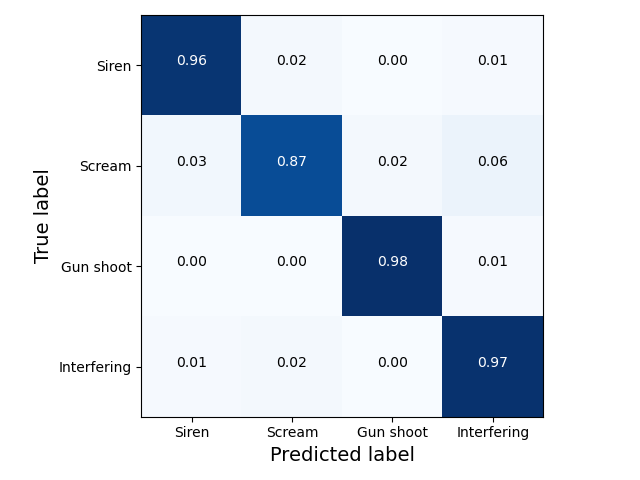}
        \label{subfig:cm_att}
    }
    \caption{\textcolor{black}{Comparison of confusion matrices obtained by the proposed method with and without attention mechanisms.}}
    \label{fig:con_matrix}
\end{figure}

\begin{figure*}[t]
\begin {center}
\includegraphics[width=0.85\textwidth]{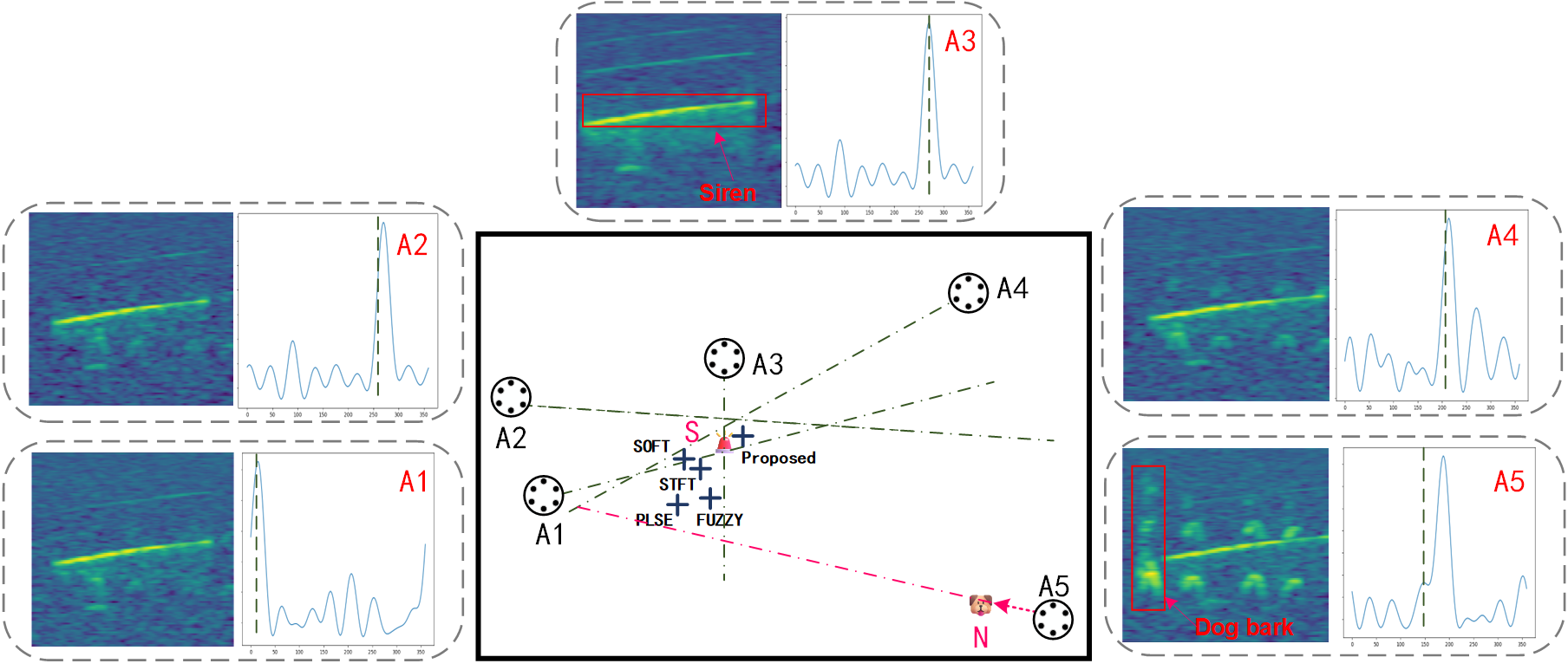}
\caption{\textcolor{black}{A typical case of the system. There are 5 array nodes, with one target sound source ($\textit{siren}$) and one interfering sound source ($\textit{dog bark}$) active simultaneously. The dashed lines on the soundmap feature represent the estimated DOA of the target sound source relative to the array nodes. A1, A2, A3, and A4 can offer relatively precise estimates of direction (green dashed line), while A5, being too close to the interfering sound source, provides inaccurate direction estimates (red dashed line). The localization results of SSL-PLSE \cite{koks2001passive} and SSL-FUZZY \cite{faraji2019sound}  are strongly impacted by A5, with RMSE values of 22.1 m and 13.3 m, respectively. By contrast, DL-based methods exhibit reduced sensitivity to A5: SSL-STFT \cite{le2019learning}  and SSL-SOFT \cite{feng2023soft} achieve RMSE of 7.3 m and 7.8 m, and our proposed method obtains a notably lower error of 3.7 m.}}
\label{fig:failure_loc}
\end {center}
\end{figure*}

\textcolor{black}{For the SEC task, we compare the performance of three baseline methods, focusing particularly on the F1 score and the FAR. 
Table \ref{tableVA} shows the experimental results: the proposed method outperforms the other three methods in terms of F1 score and FAR. 
Specifically, our method achieves significantly higher F1 scores when considering all the three classes compared to the baseline methods, with a FAR of only 0.039. }
\textcolor{black}{To explore the differences in classification performance between different sound classes, we observe that the \textit{gunshot} class consistently achieves the best classification performance, while the \textit{scream} class demonstrates the poorest performance. }
\textcolor{black}{These differences can be clearly observed in Figure \ref{fig:con_matrix}. }
\textcolor{black}{By comparing the confusion matrices of the proposed method with and without attention mechanisms, we find significant improvements introduced by the attention mechanism.
Specifically, without the attention mechanism, about 11\% of the data in the \textit{scream} class are incorrectly classified as interfering sounds, and 84\% of the \textit{scream} samples are correctly recognized.}
\textcolor{black}{Additionally, the \textit{interfering} class also exhibits relatively low accuracy (91\%), indicating notable confusion with other classes.}
\textcolor{black}{After incorporating the attention mechanism, the classification accuracy significantly improves: the accuracy of the \textit{scream} class slightly increases to 87\%, and confusion with interfering sounds reduces from 11\% to 6\%.}
\textcolor{black}{Furthermore, the accuracy of the \textit{interfering} class increases notably to 97\%, demonstrating that the attention mechanism \textcolor{black}{effectively highlights discriminative acoustic features, thereby enhancing robustness to acoustic interference.}}
\textcolor{black}{For the \textit{gunshot} class, both methods achieve high classification accuracy (96\% without attention and 98\% with attention). }
\textcolor{black}{This consistently high performance can be attributed to the distinctive impulsive acoustic characteristics of gunshots, which are less dependent on the attention mechanism.}

\textcolor{black}{To evaluate the interplay between SEC and SSL, we conducted an ablation study by modifying the proposed model. Specifically, we removed the SSL branch and tested the model's performance on the SEC task alone. 
The results indicate a decline in the classification accuracy, with the FAR increasing significantly to 0.063.
This decline can be attributed to the complexity of the target area, which typically contains two or three sound sources, only one of which is the target of interest. Without the SSL branch, the model lacks the ability to establish a correspondence between the target source's location and its associated event class. In contrast, the SSL branch helps the model to iteratively refine its predictions by learning this correspondence, thereby guiding the SEC task towards identifying the correct event class associated with the target source.}

\subsection{\textcolor{black}{Sound source localization}}

\begin{table}[b]
\centering
\caption{\textcolor{black}{SSL performance of different methods}}
\label{table_loc}
\begin{tabular}{ccccc}
\hline
\multirow{2}{*}{Metrics} & \multicolumn{4}{c}{RMSE (m) $\downarrow$}               \\ \cline{2-5} 
                         & Area 6       & Area 7        & Area 8        & Average      \\ \hline
SSL-PLSE\cite{koks2001passive}                 & 24.9         & 28.5          & 31.3          & 28.2    \\
SSL-FUZZY\cite{faraji2019sound}                & 14.7         & 15.9          & 19.4          & 16.6       \\
SSL-STFT\cite{le2019learning}                 & 8.3          & 9.7          & 10.9          &  9.6    \\
SSL-SOFT\cite{feng2023soft}                 & 7.9          & 10.2          & 13.2          &  10.4    \\
proposed (only $SSL$)    & 7.3          & 9.6          & 10.1          & 9.0        \\  
\textbf{proposed}           & \textbf{6.4} & \textbf{7.4} & \textbf{8.6} & \textbf{7.5} \\ 

\hline
\end{tabular}
\end{table}

\textcolor{black}{In Table \ref{table_loc}, we compare the performance of different SSL methods and present the localization errors across various testing areas. 
The results show that DL-based methods outperform traditional methods. 
The SSL-PLSE exhibits the poorest performance with an average localization error of $\mathrm{RMSE}=28.2$ m.
SSL-FUZZY, which optimizes upon SSL-PLSE, achieves an average error of $\mathrm{RMSE}=16.6$ m. 
SSL-FUZZY allows each array node to focus not only on a single direction but also on a directional range, effectively reducing errors caused by interfering sound sources. 
SSL-STFT uses a microphone pairwise mechanism in the neural network to learn the differences in delay and signal amplitude attenuation between microphone pairs, resulting in an average localization error of $\mathrm{RMSE}=9.6$ m.} 
\textcolor{black}{SSL-SOFT reformulates the localization task as a classification problem by employing soft-label coding. In large-scale outdoor areas, this approach can lead to noticeable quantization errors. As the monitoring region expands, these errors become more pronounced, resulting in an average localization error of $\mathrm{RMSE}=10.4$ m.}
\textcolor{black}{Our proposed method achieves an average error of $\mathrm{RMSE}=7.5$ m. 
The features we employed leverage the spatial gain of arrays to enhance useful signals and suppress unwanted interference and noise, making our method superior to those relying on STFT features. 
Figure \ref{fig:failure_loc} illustrates a typical scenario, showcasing the localization performance of different methods. 
In this scenario, five microphone arrays are distributed in the monitoring area with a target sound source ($\textit{siren}$) and an interfering source ($\textit{dog}$ $\textit{bark}$).
Most arrays provide approximate directional information of the target source. 
However, the array (A5) nearest to the interfering source, suffers from substantial interference. 
The localization results using SSL-PLSE and SSL-FUZZY are notably influenced by A5, with RMSE values of 22.1 m and 13.3 m, respectively. 
DL-based methods are less affected by A5, with SSL-STFT and SSL-SOFT achieving RMSE of 7.3 m and 7.8 m.  Our proposed method achieves 3.7 m.}

\textcolor{black}{In addition, we modify the network structure of the proposed method. 
We remove the SEC branch and only retain the SSL branch. 
Compared to the original network, removing the SEC branch reduces the proposed model's localization performance, though it still outperforms the other three baseline methods.
These results suggest that the sound event classification task contributes to improving localization accuracy. 
Consistent with what is shown in Figure \ref{fig:failure_loc}, when an interference source is present, the SEC branch assists the model in identifying nodes with features affected by interfering sound sources.}

\begin{figure}[h]
\centering
\includegraphics[scale=0.955]{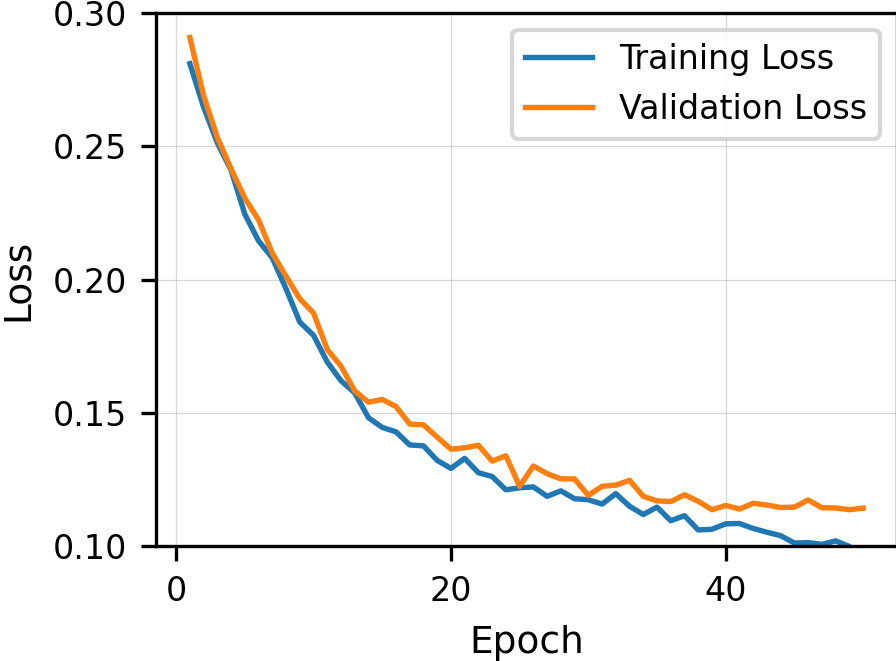} 
\caption{\textcolor{black}{Training and validation loss curves over 50 epochs, showing rapid improvement in the first 20 epochs followed by gradual convergence.}}
\label{fig:loss}
\end{figure}

\textcolor{black}{The Fig. \ref{fig:loss} shows the evolution of the training and validation loss values over 50 epochs. 
As illustrated, both training and validation losses exhibit a clear downward trend, especially during the initial 20 epochs, indicating rapid learning and significant model improvement in the early stages. 
After this period, the rate of improvement gradually slows down, and the curves stabilize around lower loss values, suggesting that the model converges effectively.}

\subsection{Sound event localization and classification}
We further compare the performance of sound event localization and classification by considering three baselines based on different SEC and SSL methods.
The first baseline employs a traditional two-stage approach, initially using SEC-CNN for sound event classification, followed by SSL-FUZZY for active sound event localization, and we refer to this method as CNN-FUZZY. 
The second baseline combines SEC-CNN and SSL-STFT in a multitask model, and we refer to this method as CNN-STFT. 
The third baseline combines SEC-CRNN and SSL-SOFT, and we refer to this method as CRNN-SOFT. 
\begin{table}[h]
\centering
\caption{\textcolor{black}{sound event localization and classification performance of different methods}}
\label{table_selc}
\begin{tabular}{ccccc}
\hline
\multirow{2}{*}{Metrics} & \multicolumn{4}{c}{$\mathrm{SELC}_{\mathrm{score}}$ $\uparrow$}                                 \\ \cline{2-5} 
                         & Area 6       & Area 7        & Area 8        & Average      \\ \hline
CNN-FUZZY                & .934         & .927          & .922          & .928       \\
CNN-STFT                & .952          & .948          & .941          &  .947    \\
CRNN-SOFT               & .955          & .939          & .932          &  .943    \\
\textbf{proposed}           & \textbf{.961} & \textbf{.958} & \textbf{.952} & \textbf{.957} \\ 
  \hline
\end{tabular}
\end{table}

In Table \ref{table_selc}, we compare the performance of different sound event localization and classification methods and present $\mathrm{SELC}_{\mathrm{score}}$ across various testing areas. 
The results indicate that the performance of the two-stage approach is inferior to that of the multitask model method. 
CNN-FUZZY has an average $\mathrm{SELC}_{\mathrm{score}}=0.928$ across the three test areas, which is lower than CNN-STFT's 0.947 and the proposed method's 0.957. 
The $\mathrm{SELC}_{\mathrm{score}}$ of each method decreases as the test area expands. 
The proposed method achieves its best $\mathrm{SELC}_{\mathrm{score}}$ in Area 6, reaching 0.961.

\subsection{Real-word experiment}

\begin{figure*}[t]
\centering
\subfigure[]{
\includegraphics[scale=0.48]{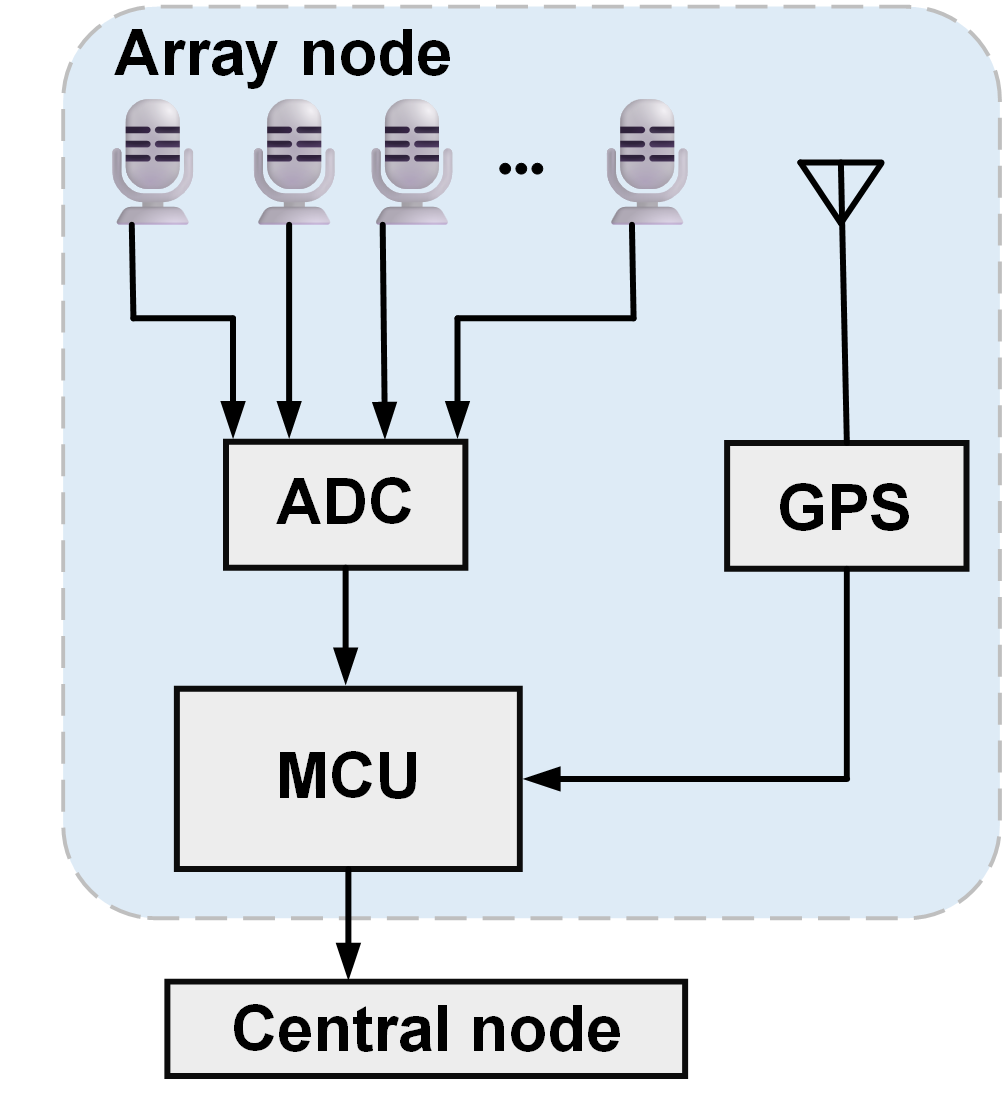} \hspace{10mm}\label{fig:array} 
}
\subfigure[]{
\includegraphics[scale=0.48]{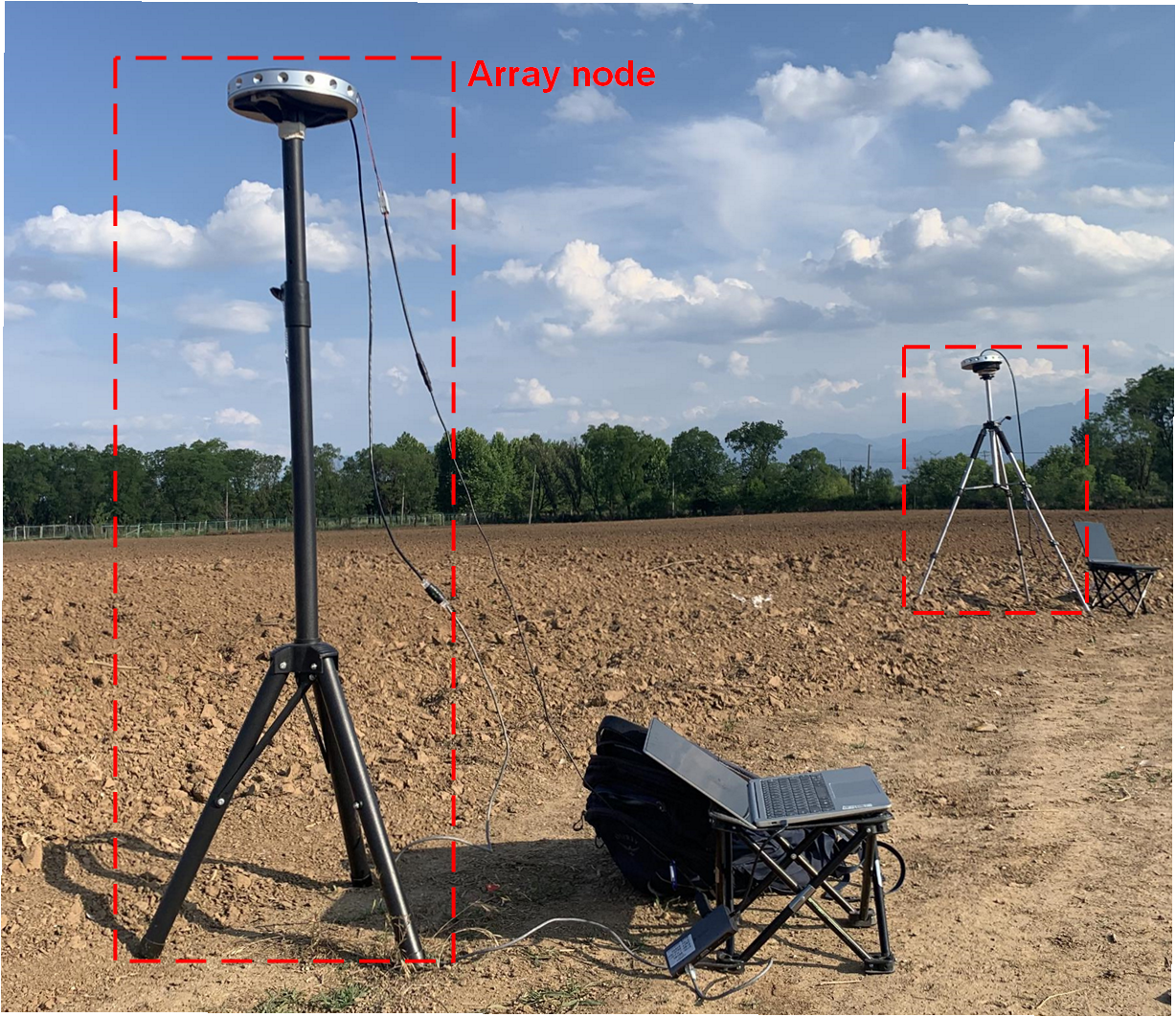} \label{fig:setup} 
}
\quad
\caption{\textcolor{black}{Deployment and schematic representation of the WASN for outdoor acoustic monitoring. (a) Structure of an array node composed of multiple microphones. (b) Real-world setup of array nodes in an open outdoor environment.}}
\end{figure*}

We validate our proposed method using recordings sampled in a real-world setting, as illustrated in Fig.~\ref{fig:setup}. 
The experimental data are sampled in an urban park. 
We designate a specific area within the park with dimensions of 100~m $\times$ 80~m. 
The recordings are acquired from five circular arrays, each equipped with 8 microphone elements, with a radius of 11 cm for each array. 
The sampling rate of the array node is set at 8 KHz. 
We use two Bluetooth speakers as sound sources, and all array nodes are placed at a height of 1.5 m. 

\textcolor{black}{In the proposed WASN system, each array node is not only responsible for collecting multi-channel acoustic signals but also for performing on-board edge-level data preprocessing and feature extraction. 
After the computation, the time-stamped results are transmitted to the central node for higher-level fusion and analysis. 
Fig. \ref{fig:array} illustrates the overall hardware architecture of each array, including the functional modules and the data flow.}

\textcolor{black}{Each array node is built around an STM32F4 microcontroller unit (MCU) based on the ARM Cortex-M4 architecture, as illustrated in Fig.~\ref{fig:hw_1}. 
The STM32F4 runs at a maximum clock frequency of 168 MHz, provides a floating-point unit, and supports multiple peripheral interfaces (e.g., UART, SPI), thereby meeting the multi-channel data acquisition, digital signal processing, and high-speed data transmission requirements. 
For acoustic signal acquisition, each array node is equipped with multiple microphones, and an external analog-to-digital converter (ADC) chip converts the analog signals into multi-channel digital data. 
In this system, the AD7606 is chosen.
When the system is operational, multi-channel microphone signals are continuously fed into the AD7606. 
The digital audio data is then transferred to the MCUs’ on-chip memory via the Direct Memory Access (DMA) channel. 
To achieve seamless high-speed data reading and processing, the system enables the DMA in double buffer mode, such that while one buffer is being filled with new data, the other buffer can be simultaneously accessed for signal processing. 
This design effectively avoids data overwriting or loss while ensuring real-time processing capability.}

\textcolor{black}{For local feature extraction at the edge side, the STM32F4 utilizes the CMSIS-DSP library \cite{wickert2015using} provided by ARM. 
This library includes commonly used digital signal processing functions (e.g., filtering, transformation, and Fourier analysis), allowing preprocessing of the sampled multi-channel audio \textcolor{black}{frames and extraction of both Soundmap and GTGram features.} }
\textcolor{black}{Consequently, the computational burden and data transmission load on the central node are considerably reduced. 
Field tests show that generating Soundmap and GTGram features from one second of multi-channel audio data requires only 380 ms on the STM32F4, meeting the system’s real-time requirements.
To achieve consistent time referencing }\textcolor{black}{and data synchronization across multiple nodes, each array node is equipped with a GPS module. }
\textcolor{black}{The GPS provides \textcolor{black}{a highly accurate Pulse-Per-Second (PPS) signal and standard time information (commonly in NMEA format, reporting year, month, day, hour, minute, and second, etc.).} }
\textcolor{black}{Specifically, the PPS pin is connected to the STM32F4 external interrupt pin configured for rising-edge triggering. 
Whenever a pulse arrives each second, the MCU generates an external interrupt and, in the corresponding interrupt service routine, cooperates with the DMA to enable or reset operations. 
This ensures that the audio frames stored in the on-chip memory are aligned with the arrival of the PPS signal, as illustrated in Fig.~\ref{fig:hw_2}. }
\textcolor{black}{Meanwhile, the MCU periodically reads the standard timestamp output by the GPS through a serial port, associating it with the corresponding audio data or features, thus achieving precise time synchronization.
After feature extraction and time alignment, each array node transmits the acoustic features with timestamps to a network chip (W5500 in this example) via the SPI interface. 
The W5500 establishes a TCP connection with the central node, delivering the data in real time.}

\textcolor{black}{During the experiment, we randomly distribute the locations of the sound sources and array nodes. 
We also set up 30 position arrangements of sources and recorded 60 seconds of audio for each arrangement and each sound event class. 
We fine-tune the model by using 20 position arrangements. 
Furthermore, we augment the training data by varying the input order of the indices of the arrays when we generate features. }

\begin{figure*}[t]
\centering
\subfigure[]{
\includegraphics[scale=0.46]{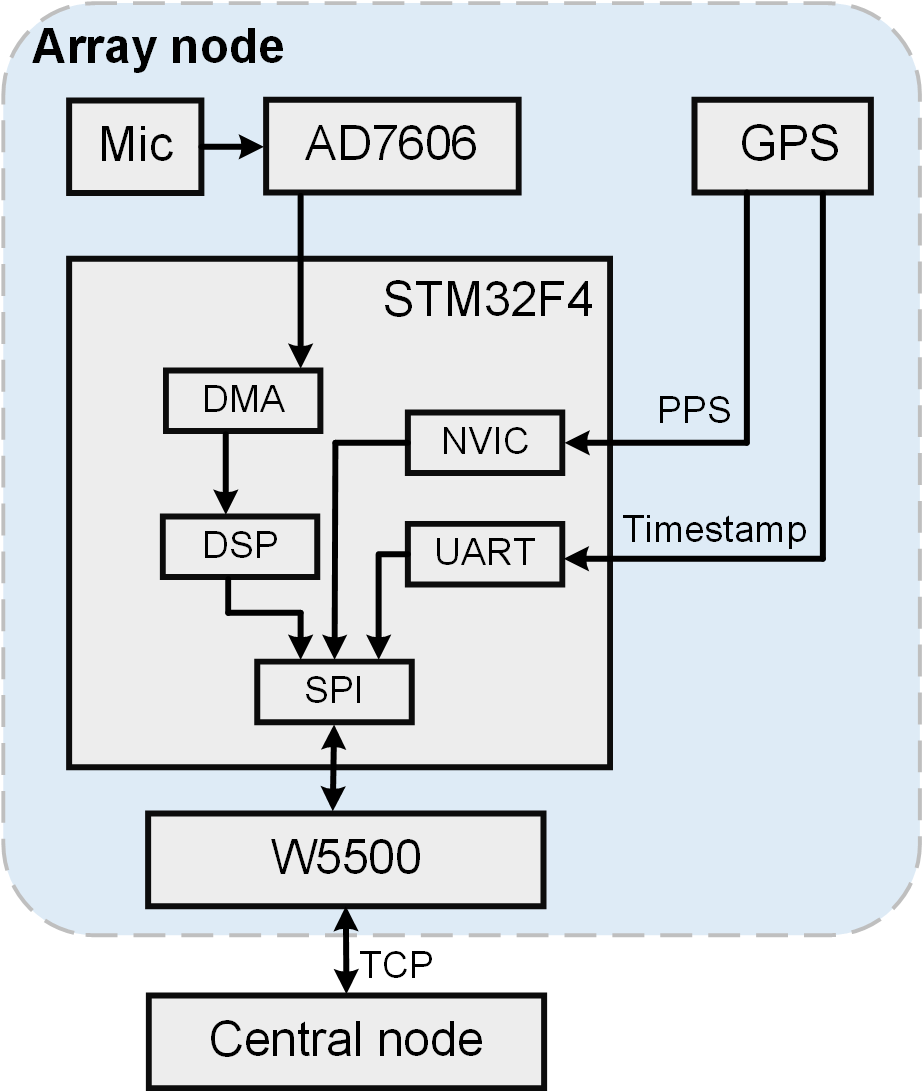} \hspace{10mm}\label{fig:hw_1} 
}
\subfigure[]{
\includegraphics[scale=0.46]{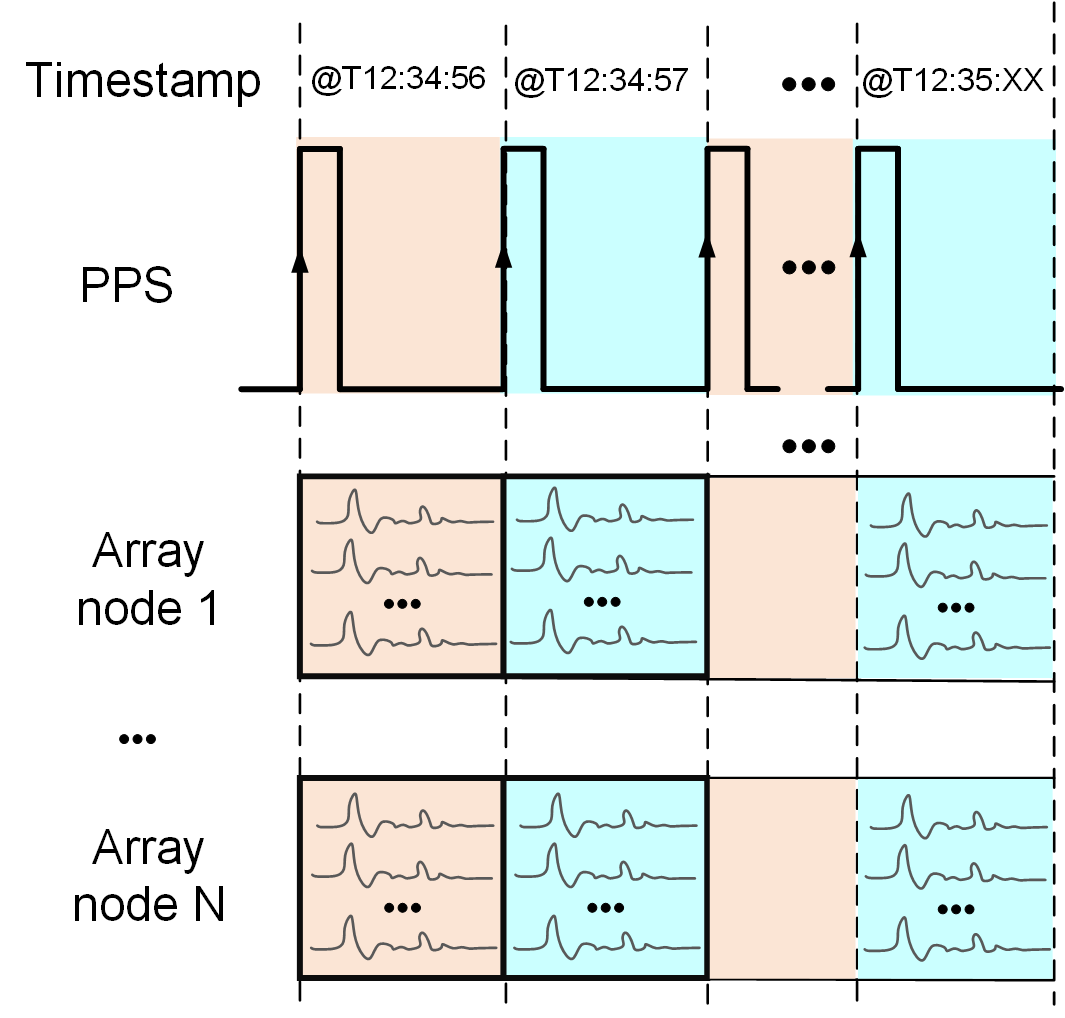} \label{fig:hw_2} 
}
\quad
\caption{\textcolor{black}{Architecture of an array node and synchronization mechanism within the WASN. (a) Internal structure of the array node, including microphone, analog-to-digital converter, microcontroller unit, and GPS module for synchronization. (b) Synchronization process illustration, showing precise timestamps obtained through GPS signals and PPS synchronization pulses across multiple nodes.}}
\end{figure*}

\textcolor{black}{We randomly distributed both the array nodes and the sound sources, ensuring a diverse coverage of potential positions. 
Specifically, we arranged 30 distinct source–array layouts and recorded 60 seconds of continuous audio for each arrangement and each sound event class. 
Of these configurations, 20 were used for fine-tuning, and the remaining 10 were reserved for evaluation to simulate new unseen scenarios.
To increase the robustness of our model and enhance generalization, we employed data augmentation strategies. 
First, we varied the activation sequence of the arrays when generating the spatio-temporal features, effectively randomizing the input channel order. 
This helps the model learn invariant representations regardless of how arrays are indexed or ordered. 
Second, we introduced slight perturbations to the source positions and background noise levels in software by overlaying low-level ambient recordings from other urban areas, improving the system’s ability to handle unanticipated environmental sounds. 
Third, we used an early-stopping criterion and regularization techniques to avoid overfitting to the specific acoustic characteristics of our initial training subset.}

\begin{table}[h]
\centering
\caption{\textcolor{black}{SEC performance in real-world scenarios}}
\label{table:exp_sed}
\begin{tabular}{ccccccc}
\hline
\multirow{2}{*}{Metrics} & \multicolumn{3}{c}{$\mathrm{F1}$ score $\uparrow$} & \multirow{2}{*}{$\mathrm{FAR}_{20}$ $\downarrow$} \\ 
\cline{2-4}  
                         & siren          & scream         & gunshot           &                 \\ \hline
SEC-CNN\cite{mushtaq2020environmental}                  & .918          & .832          & .943                   & .103        \\
SEC-UNET\cite{marchegiani2022listening}                  & .927          & .875          & .962                   & .112       \\
SEC-CRNN\cite{bansal2024robust}                  & .932          & .869          & .970                   & .101       \\
\textbf{proposed}           & \textbf{.955} & \textbf{.889} & \textbf{.976} & \textbf{.067}      
               
\\\hline
\end{tabular}
\end{table}

\textcolor{black}{For the SEC task, Table \ref{table:exp_sed} presents the experimental results. 
We first conduct tests on the false alarm rates of each algorithm. 
The WASN system only collects environmental sounds in the park, and the $\mathrm{FAR}_{20}$ denotes the ratio of environmental sounds classified as active class within a 20-minute interval. 
Compared to simulated experiments, the performance of all methods experiences a decline when tested with data recorded in real-world scenarios. 
However, our proposed method still outperforms the other two baseline methods, achieving F1 scores of 0.955, 0.889, and 0.976 for the three sound event classes, respectively, with a false alarm rate of 0.067.
Further analysis reveals that the primary reasons for false alarms are children playing and street music near the site, and vehicles honking approximately 300 m away on the road. 
The sounds of percussion instruments in street music resemble gunshot sounds, both being non-stationary impulsive signals. }
\begin{table}[h]
\centering
\caption{\textcolor{black}{SSL performance in real-world scenarios}}
\label{table:loc}
\resizebox{\linewidth}{!}{
\begin{tabular}{cccccc}
\hline
Methods &  PLSE\cite{koks2001passive}  & FUZZY\cite{faraji2019sound}  & STFT\cite{le2019learning} & SOFT\cite{feng2023soft}  &  \textbf{proposed} \\  \hline
RMSE (m)     & 11.2 & 9.7 & 5.8 & 6.1 & \textbf{4.5} \\
\hline
\end{tabular}
}
\end{table}

\textcolor{black}{For the SSL task, Table \ref{table:loc} presents the experimental results. 
Compared to simulated experiments, all SSL methods experienced a decrease in performance when tested with data recorded in real-world scenarios. 
However, our proposed method still outperforms the other three baseline methods, with $\mathrm{RMSE}=4.5$ m. }

\begin{table}[h]
\centering
\caption{\textcolor{black}{sound event localization and classification performance in real-world scenarios}}
\label{table:loc}
\resizebox{\linewidth}{!}{
\begin{tabular}{ccccc}
\hline
Methods           &   CNN-FUZZY  & CNN-STFT & CRNN-SOFT &  \textbf{proposed} \\  \hline
$\mathrm{SELC}_{\mathrm{score}}$    & .902          & .919   & .922       & \textbf{.946} \\
\hline
\end{tabular}
}
\end{table}

\textcolor{black}{Then we validate the overall performance of the proposed method. 
By comparing CNN-FUZZY, CNN-STFT, and CRNN-SOFT, we can conclude that our proposed method shows the best performance in real-world experiments, with $\mathrm{SELC}_{\mathrm{score}}= 0.946$. 
Further analysis reveals that the primary cause of the performance decline is the complexity of the real environment.
Although the park is an open area, factors such as trees, benches, and pedestrians disrupt sound propagation.
The robustness of the model is challenged by the reflection and absorption of sound waves caused by these objects.}

\section{CONCLUSION}

\textcolor{black}{This paper proposed a DL-based method for the localization and classification of sound events using WASNs.
We employ multiple microphone arrays, each equipped with several microphone sensors, to sample and process acoustic signals.
We introduce a novel feature that integrates information from the frequency, temporal, and spatial domains.
In addition, we present a network architecture that uses attention mechanisms to learn channel-wise relationships and temporal dependencies within the acoustic features.
The experimental results demonstrate the superiority of our proposed method over baseline approaches.
Specifically, our method achieve the highest F1 score and FAR in the SEC task, and the lowest RMSE in the SSL task.
Further experiments confirm the mutual enhancement of learning capabilities between the SEC and SSL tasks.
The visualization of localization errors highlight the robust SSL performance of our proposed method in complex environments.
Finally, the efficacy of the proposed method is further validated by real-world experiments.}
\textcolor{black}{Building on this robust foundation, future work will aim to enhance the physical realism of our simulation framework. 
We plan to investigate the integration of more advanced modeling techniques to account for complex, time-varying environmental factors such as wind, temperature gradients, and turbulence, further improving the model's resilience in diverse real-world conditions.}

\bibliographystyle{elsarticle-num} 
\textcolor{black}{\bibliography{ref}}

\end{document}